\documentclass[pra,twocolumn,showpacs,floatfix]{revtex4}
\usepackage{graphicx}
\begin{document}

\title{Persistent currents in dipolar Bose-Einstein condensates 
confined in annular potentials}
\author{F. Malet$^1$, G. M. Kavoulakis$^2$, and S. M. Reimann$^1$}
\affiliation{$^1$Mathematical Physics, Lund Institute of Technology,
P.O. Box 118, SE-22100 Lund, Sweden \\
$^2$Technological Educational Institute of Crete, P.O. Box 1939, 
GR-71004, Heraklion, Greece}
\date{\today}

\begin{abstract}
We consider a dipolar Bose-Einstein condensate confined in 
an annular potential, with all the dipoles being aligned along 
some arbitrary direction. In addition to the dipole-dipole 
interaction, we also assume a zero-range hard-core potential. 
We investigate the stability of the system against collapse, 
as well as the stability of persistent currents as a function 
of the orientation of the dipoles and of the strength of the 
hard-core interaction. 
\end{abstract}


\maketitle

\section{Introduction}

Ultracold gases of bosonic and fermionic atoms have shown 
numerous fascinating effects. In some relatively recent 
experiments, it has become possible to create Bose-Einstein 
condensates of dipolar atoms \cite{Grie, Fatt, Veng, Lu}, 
and also to confine ultracold polar molecules \cite{Cosp, Sosp}. 
Dipolar gases may give rise to novel physics as compared to 
non-dipolar ones: unlike the usual hard-core potential, which
describes the effective atom-atom interaction of non-dipolar 
atoms, the dipole-dipole interaction is anisotropic, nonlocal
and it may be partly attractive and partly repulsive. For 
example, even when all the dipoles are aligned (under, e.g., 
the action of an electric or magnetic field, as we consider 
in the present study), the sign and the strength of the 
interaction may vary spatially, and can be tuned externally 
by changing the orientation of the applied field 
\cite{Bar, Lah}. 

In some other recent experiments, toroidal trapping 
potentials have been realized experimentally 
\cite{Gup, Ols, Ryu, Hen, Les}, where Bose-Einstein
condensates of non-dipolar atoms were observed to
support persistent currents \cite{Ryu}. The realization 
of a dipolar Bose-Einstein condensate trapped in 
toroidal/annular potentials should therefore be feasible 
experimentally, as well. A condensate in such a topologically 
non-trivial confining potential is an ideal system for the 
investigation of various interesting phenomena, including 
superfluid properties, nonlinear effects, etc.

Theoretical studies of dipolar atoms trapped in toroidal 
potentials have been performed recently \cite{Aba10,Aba2,Zol11}. 
In Ref.\,\cite{Aba10}, Abad {\it et al.}\,\,considered 
a three-dimensional trapping geometry and used the 
Gross-Pitaevskii mean-field approximation to investigate 
the ground-state and the rotational properties of a dipolar 
condensate for a fixed orientation of the dipoles, which 
was chosen to be on the plane of motion of the atoms (say, 
the $x$-$y$ plane). It was shown that if the dipoles are 
pointing along, e.g., the $x$ axis, there is a high 
concentration of them around the two points with coordinates 
$(x,y,z)=(0,\pm R,0)$, with $R$ being the mean radius of 
the torus, since around these two points the dipoles are 
predominantly oriented head-to-tail and therefore the 
interaction is mostly attractive. As we see below, the 
dipolar energy has two degenerate minima at these two 
points. As a result, as the strength of the dipolar 
interaction increases -- with all the other parameters 
kept fixed -- the density shows first two peaks around 
them. Eventually this two-fold symmetry is broken and 
the cloud concentrates around only one of them. If the 
dipolar interaction becomes strong enough, the gas 
collapses. In the same study it was also found that the 
gas supports metastable currents. In a more recent paper 
\cite{Aba2} the same authors demonstrated that this system 
exhibits Josephson oscillations and macroscopic quantum 
self-trapping. Finally, in Ref.\,\cite{Zol11}, Z\"ollner 
{\it et al.} considered quasi-one-dimensional motion of 
aligned dipolar atoms along a toroidal trap, assuming a 
very tight confinement in the transverse direction, and 
studied the few-body regime with the method of numerical 
diagonalization of the many-body Hamiltonian.

In the present study we consider a dipolar Bose-Einstein 
condensate confined in a (quasi-two-dimensional) annular 
trap, and assume that all the dipoles are aligned along 
some direction, forming an angle $\Theta$ with their plane 
of motion. In addition to the dipolar interaction, we consider 
a zero-range contact potential with strength $g$. Using the 
mean-field, Gross-Pitaevskii approximation we determine the 
absolute and/or local minima of the energy of the system for 
several values of $g$ and $\Theta$. We identify a collapsed 
phase, in which the system cannot support itself, and a stable 
one. Within the stable phase we also find a subphase in which 
the gas supports metastable, persistent currents. 

Our results show that the boundaries between the different 
phases have a sinusoidal dependence on the angle $\Theta$, 
in agreement with a simple variational model that we present. 
The atomic density distribution that we find resembles 
that of the study by Abad {\it et al.} When the dipole moment 
is perpendicular to the plane of motion of the atoms the 
density is axially symmetric. As the angle is tilted, the 
density develops two maxima, and eventually the cloud 
concentrates around only one of these two maxima. We also 
find that, as observed in Ref. \cite{Aba10} for $\Theta=0$,
even when the particle density is not circularly symmetric, 
the system may support persistent currents, provided 
that the atoms are not concentrated around a single density 
maximum. This behaviour is in a sense reminiscent of the case 
of purely contact interactions, where, if the ground-state 
density localizes, the system undergoes solid-body rotation 
and there is no metastability. 

In what follows we first present our model in Sec.\,II. 
In Sec.\,III we investigate the stability of the system 
against collapse, and in Sec.\,IV the stability of persistent 
currents. Finally, in Sec.\,V we present our conclusions. 

\section{Model}

We consider a dipolar Bose-Einstein condensate at zero 
temperature, confined in an axially-symmetric potential
\begin{eqnarray}
V({\bf r}) = V_r({\bf r}_{\perp}) + V_z(z) =
\frac 1 2 M \omega^2 (r_{\perp} - R)^2 
+ \frac 1 2 M \omega_z^2 z^2,
\nonumber \\
\end{eqnarray}
where the $z$ direction is chosen to be the symmetry axis of 
the potential, and ${\bf r}_{\perp}$ is the position vector 
on the $x$-$y$ plane. Also, $\omega_z$ and $\omega$ are the 
frequencies of the confinement along the $z$ axis and along 
the direction perpendicular to it. We choose $\omega_z/\omega$ 
equal to 100, in order for $\hbar \omega_z$ to be much larger 
than all the other energy scales in the problem, in which case 
the motion is quasi-two-dimensional. Finally, we consider 
$R/a_0 = 4$, where $a_0 = [\hbar/(M \omega)]^{1/2}$ is the 
oscillator length corresponding to $\omega$, and $M$ is the 
atomic mass. Under these conditions $V({\bf r})$ describes 
an annular potential with mean radius $R$ and width $\approx 
a_0$, as shown schematically in Fig.\,1.

\begin{figure}[t]
\centerline{\includegraphics[width=6cm,clip]{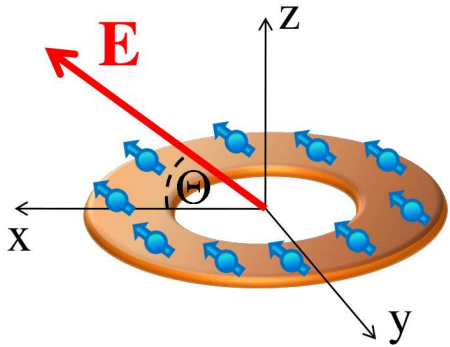}}
\caption{(Color online)
Schematic illustration of the quasi-two-dimensional potential. 
A very tight confinement is assumed along the $z$ axis, while 
on the $x$-$y$ plane the atoms are assumed to move in an annulus 
of mean radius $R$ and width $\approx a_0$. The dipoles are 
oriented on the $x$-$z$ plane forming an angle $\Theta$ with 
the $x$ axis due to the action of an external (magnetic or 
electric) field ${\bf E}$.} 
\label{fig1}
\end{figure}

Turning to the interactions, we consider both the usual  
contact potential and the dipole-dipole interaction,
\begin{eqnarray}
V_{\rm int, 3D}({\bf r},{\bf r}') = U_0 \delta({\bf r}-{\bf r}')+
D^2 \frac {1-3\cos^{2} \theta_{rd}} {|{\bf r} - {\bf r}'|^{3}},
\label{intef}
\end{eqnarray}
where $U_0$ is the matrix element for zero-energy elastic
atom-atom collisions. When the atoms have an electric dipole 
moment $d$, $D^2 = d^2/(4 \pi \epsilon_0)$, where $\epsilon_0$ 
is the permittivity of the vacuum, while when the atoms have a 
magnetic moment $\mu$, $D^2 = \mu_0 \mu^2/(4 \pi)$, where $\mu_0$ 
is the permeability of the vacuum. Finally, $\theta_{rd}$ is the 
angle between the dipole moment and the relative position vector 
${\bf r} - {\bf r}'$ of two dipoles. 

The assumption of strong confinement along the $z$ axis
implies that the motion of the atoms is frozen along this 
direction. This allows us to make an ansatz for the order 
parameter of the form $\Psi({\bf r})=Z(z)\Phi({\bf r}_{\perp})$, 
with $Z(z) = e^{-z^2/2a_z^2}/(\pi a_z^2)^{1/4}$ being the ground 
state of the harmonic oscillator with frequency $\omega_z$, and 
$a_z = [\hbar/(M \omega_z)]^{1/2}$. With this ansatz, one may 
integrate the dipole-dipole interaction along the $z$ direction 
and derive an effectively two-dimensional dipolar potential. 
After performing the integration, this potential takes the form 
\cite{Cre10}
\begin{eqnarray}
  V_{\rm eff}({\bf r}_{\bot})=\frac{D^{2}}{\sqrt{8 \pi}}
  \frac{e^{\tilde r_{\perp}^2}}{a_z^3}\biggl\{(2+4 \tilde r_{\perp}^2)
  K_{0}(\tilde r_{\perp}^2)-4 \tilde r_{\perp}^2 
  K_{1}(\tilde r_{\perp}^2)
\nonumber\\
 +\cos^{2}\Theta \biggl[-(3+4 \tilde r_{\perp}^2) 
 K_{0}(\tilde r_{\perp}^2)+(1+4 \tilde r_{\perp}^2) 
 K_{1}(\tilde r_{\perp}^2)\biggl]
\nonumber\\
 +2\cos^{2}\Theta\cos^{2}\phi\biggl[- 2 \tilde r_{\perp}^2 
 K_{0}(\tilde r_{\perp}^2)+(2 \tilde r_{\perp}^2-1)
 K_{1}(\tilde r_{\perp}^2)
\biggl]\biggl\},
\label{eqVdipeff}
\end{eqnarray}
where $\phi$ is the angle in cylindrical polar coordinates, 
$\tilde r_{\perp} \equiv r_{\bot}/(2 a_z)$, and $K_0$ and 
$K_1$ are the zeroth-order and first-order modified Bessel
functions of the second kind. 

Integrating also the contact interaction over the $z$ direction
we find that the total effective interaction is given by
\begin{eqnarray}
V_{\rm int, 2D}({\bf r}_{\bot}) = g \delta({\bf r}_{\bot}) + 
V_{\rm eff}({\bf r}_{\bot}),
\label{totint}
\end{eqnarray} 
where $g \equiv U_0 \int |\phi_0(z)|^4 dz = U_0/(\sqrt{2 \pi} 
a_z)$. Therefore, $\Phi({\bf r}_{\perp})$ satisfies the 
Gross-Pitaevskii-like equation 
\begin{eqnarray}
\biggl[- \frac {\hbar^2 \nabla_{\bot}^2}{2 M} + 
V_r({\bf r}_{\bot}) 
+ V_{\rm dip}({\bf r}_{\bot}) 
+ g |{\Phi({\bf r}_{\bot})}|^2 \biggl]  
\Phi({\bf r}_{\bot}) = 
\nonumber \\ = \mu \Phi({\bf r}_{\bot}),
\label{gpe}
\end{eqnarray}
where $\mu$ is the chemical potential, and
\begin{equation}
V_{\rm dip}({\bf r}_{\bot}) = \int 
V_{\rm eff}({\bf r}_{\bot}-{\bf r}_{\bot}^{\prime}) \,
\left|\Phi({\bf r}_{\bot}^{\prime})\right|^2\,
d{\bf r}_{\bot}^{\prime}
\end{equation}
is the effective dipolar interaction potential. In order to 
solve Eq.\,(\ref{gpe}), we have employed a fourth-order 
split-step Fourier method within an imaginary-time propagation 
approach \cite{Chi05}. The method requires the selection of 
an initial state, which is then propagated in imaginary time 
until a (local or absolute) minimum of the energy is reached 
and numerical convergence is achieved. In what follows we fix
the value of the dipolar length $a_{dd} \equiv M D^2/(3 \hbar^2)$, 
to the value of $a_{dd}/a_0 = 4/3$, and vary $\Theta$ and $g$.

\section{Stability of the ground state}

We first investigate the ground state of the system. Since the 
dipolar interaction may be partly attractive, and we also allow
for negative values of the contact interaction, we start with 
the stability of the ground state against collapse. For this 
purpose, we choose the initial guess for $\Phi({\bf r}_{\perp})$ 
to be proportional to $\exp[-(r_{\perp}-R)^2/2a_0^2]$, i.e., an 
axially symmetric Gaussian, which is peaked around $r_{\perp}=R$,
with a width determined by $a_0$. For a given $\Theta$, we solve 
Eq.\,(\ref{gpe}) for different values of $g$, identifying the 
different phases, and then we plot the corresponding phase 
diagram $g = g(\Theta)$, which is shown in Fig. 2.

The boundary separating the collapsed from the stable phase 
can be fitted rather well by the following sinusoidal 
function
\begin{equation}
g_{c,{\rm stab}}/g_0 \approx 11.0 \cos(2 \Theta) - 3.5,
\label{gstab}
\end{equation}
with $g_0 \equiv D^2/(4 a_z)$, which is represented by the 
dashed curve in Fig.\,2. As expected, the minimum value of 
$g$ necessary to prevent the collapse decreases as the 
orientation angle of the dipoles $\Theta$ increases, since 
in this case the dipole-dipole interaction becomes 
increasingly repulsive. 

\begin{figure}[t]
\centerline{\includegraphics[width=9cm,clip]{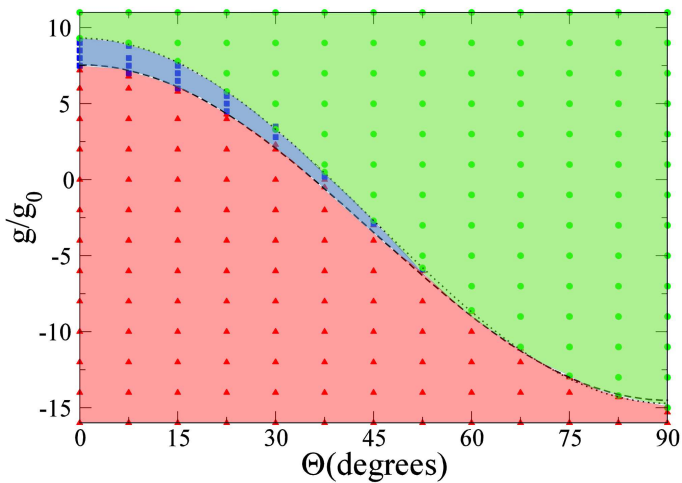}}
\caption{(Color online)
Phase diagram with the orientation of the dipoles $\Theta$ in 
the horizontal axis and the strength of the contact potential  
$g$ in the vertical one. The three phases discussed in the 
text are presented: the stable phase where there exist 
metastable states (green circles), the stable phase that does
not support persistent currents (blue squares), and the
collapsed phase (red triangles). The dashed and dotted lines 
represent the fitting functions of Eqs.\,(\ref{gstab}) and 
(\ref{gmet}), respectively.}
\label{fig2}
\end{figure}

In order to get a better understanding of the above results, 
it is instructive to consider a variational order parameter 
of the form $\Psi({\bf r}) = Z(z) \Phi_{\perp}(s_{\perp}) 
\Phi_l(s_l)$, where all the three functions are assumed to 
be Gaussian with widths determined by the oscillator length
$a_z$, and the variational parameters $d_{\perp}$ and $d_l$
(as discussed in detail in the Appendix). Here $s_{\perp}$ 
is the coordinate that corresponds to the transverse 
direction and is measured from the minimum of the annular
potential. Also, $s_l$ is the coordinate that corresponds to 
the longitudinal direction and is measured from one of the 
two minima of the dipolar interaction. As shown in the 
Appendix, we first integrate the dipole-dipole interaction 
entering Eq.\,(\ref{intef}) along the vertical and transverse 
directions, assuming $a_z \ll d_{\perp}$. Then, if we 
consider two dipoles located at the positions $s_l$ and 
$s_l'$ along the annulus, for small values of $|s_r| 
\equiv |s_l-s_l'|$, we obtain an effectively 
one-dimensional dipolar potential, which is given by 
\begin{eqnarray}
V_d(s_l, s_l') \approx \frac {\sqrt{8 \pi}} {3 \pi}
\frac {N D^2} {d_{\perp} s_r^2}
[1 - 3 \sin^2 (S/R) \cos^2 \Theta].
\label{vdip1d}
\end{eqnarray}  
As one can see, $V_d(s_l, s_l')$ separates in the relative
coordinate $s_r=s_l-s_l'$ and the center-of-mass coordinate
$S=(s_l + s_l')/2$, as in the case of quasi-one-dimensional 
motion in a toroidal trapping potential \cite{Meystre}; see 
also Refs.\,\cite{Santos,Frank}. In Eq.\,(\ref{vdip1d}), 
the term that depends on the center-of-mass coordinate 
$S$ is $\propto 1 - 3 \sin^2 (S/R) \cos^2 \Theta$, and 
has two degenerate local minima, at the points with 
coordinates $(x,y)=(0,\pm R)$, if $\Theta \neq 90^\circ$, 
as expected from the arguments presented earlier.

Integrating also in the longitudinal direction, under
the assumption of a Gaussian profile, we find that the 
energy of the gas is 
\begin{eqnarray}
E &\approx& \left( \frac {\hbar^2} {2 M a_z^2} +
\frac 1 2 M \omega_z^2 a_z^2 \right) +
\left( \frac {\hbar^2} {2 M d_{\perp}^2} +
\frac 1 2 M \omega_{\perp}^2 d_{\perp}^2 \right) +
\nonumber \\
&+& \frac {\hbar^2} {2 M d_l^2}
+ \frac {N g} {4 \pi} \frac 1 {d_l d_{\perp}} +
 \frac {4} {3 \pi}
\frac {N D^2} {d_l d_{\perp} a_z}
\left(1 - \frac 3 2 \cos^2 \Theta \right).
\nonumber \\
\label{trial}
\end{eqnarray}
In evaluating the last term we have introduced a cut-off 
length in the integration over $s_r$ set by the oscillator 
length $a_z$ in the $z$ direction. Also, the factor 3/2 in 
the same term comes from averaging the function $\sin^2(S/R)$ 
[that enters $V_{\rm CM}(S)$] along the longitudinal 
direction. This factor is accurate when $d_l$ is comparable 
to $R$, or in other words when the density in the longitudinal 
direction is either homogeneous, or close to homogeneous, as 
compared to the spatial variation of $V_{\rm CM}$, which is 
of the order of $R$.

Four important conclusions follow from Eq.\,(\ref{trial}):

(i) First of all, this problem resembles that of a non-dipolar 
system in two dimensions interacting via an effective contact 
potential with strength $g_{\rm eff} = g + {\cal C} D^2 (1-3 
\cos^2 \Theta/2)/a_z$, where ${\cal C}$ is a dimensionless 
constant of order unity.

(ii) If the motion in the transverse direction is frozen, 
the variational length $d_{\perp}$ may be set equal to the 
oscillator length $a_0$, and $d_l$ is the only free parameter. 
The motion is then quasi-one-dimensional and, for small 
values of $d_l$, the kinetic energy, which scales as 
$d_l^{-2}$, dominates over the interaction energy, which 
scales as $d_l^{-1}$. Thus, in this case the total energy is 
always bounded and there is no collapse. On the other hand, 
if the motion in the transverse direction is not frozen (i.e., 
as the annulus becomes wider), the system is not necessarily 
stable against collapse: in this case, the interaction energy 
scales as $(d_l d_{\perp})^{-1}$, and the kinetic energy scales 
as $d_l^{-2}$ and $d_{\perp}^{-2}$ along the longitudinal
and the transverse directions, respectively. Therefore, the 
energy is not necessarily bounded and the system may collapse, 
under the conditions that we investigate below.

(iii) Provided that the system is stable, the energy has a local 
minimum that determines the width of the cloud in the transverse 
and longitudinal directions.

(iv) Finally, from Eq.\,(\ref{trial}) one may also deduce the
functional form of the minimum value of $g(\Theta)$ that is 
necessary for the stability of the gas against collapse. Since 
$a_z$ is chosen to be small, the last two terms in Eq.\,(\ref{trial}) 
are the dominant ones. As a result, the critical value of $g$ is 
given by the approximate expression
\begin{eqnarray}
\frac {g_{c,{\rm stab}}} {g_0} \approx \frac {16} 3
[3 \cos (2 \Theta) - 1],
\label{pb}
\end{eqnarray}
which is in rather good agreement with Eq.\,(\ref{gstab}).

\section{Persistent currents}

We turn now to the second main question of the present 
study, which is the stability of persistent currents. To
answer this question we choose the initial state to be 
proportional to $e^{i m \phi}\exp[-(r_{\perp}-R)^2/
2a_{\perp}^2]$, which has the same density distribution 
as before, but now with $m$ units of circulation. For a 
given value of $g$ and $\Theta$ (in the part of the phase 
diagram where the ground state is stable), the final state 
may or may not preserve the circulation, indicating the 
support or not of persistent currents for the particular 
choice of parameters. These two situations are represented 
in Fig.\,2, with green circles and blue squares, respectively, 
for one unit of circulation, $m=1$. The boundary between these 
two phases is given approximately by
\begin{equation}
{g_{c,{\rm met}}}/{g_0} \approx 12.0 \cos(2 \Theta) - 2.7
\label{gmet}
\end{equation}
and is represented in Fig.\,2 by the dotted curve. In 
analogy to the case of non-dipolar atoms \cite{Bar10}, 
we have observed metastable states with up to four units 
of circulation, $m=4$ with the minimum value of $g$ 
necessary for the stability of the currents increasing 
with $m$.

Again, one can get some insight into this result from 
the variational model presented in the previous section. 
For a Bose-Einstein condensate with purely contact 
interactions that is confined in a ring potential, the 
critical value of $U_0$ for the existence of persistent 
currents with one unit of circulation satisfies the equation 
$n_0 U_0/E_{\rm kin} = 3/2$, where $E_{\rm kin}= \hbar^2/
(2 M R^2)$ and $n_0$ is the density of the gas (homogeneously
distributed) \cite{U, GMK}. Using this result for the effective 
coupling constant $g_{\rm eff}$, one gets
\begin{eqnarray}
\frac {g_{c,{\rm met}}} {g_0} \approx 
\frac {16} 3 [3 \cos (2 \Theta) - 1]
- 3 \sqrt{2 \pi} \, \frac {\hbar^2} {M R^2}
\frac {R a_z a_{\perp}} {N D^2}.
\end{eqnarray}
The above formula differs from Eq.\,(\ref{pb}) in the 
last term. However, this term is small, and therefore 
the variational model implies that $g_{c,{\rm met}}$ 
varies sinusoidally with $\Theta$, and also that the 
two phase boundaries are close to each other, in agreement
with the numerical results.

\begin{figure}[t]
\centerline{\includegraphics[width=9cm,clip]{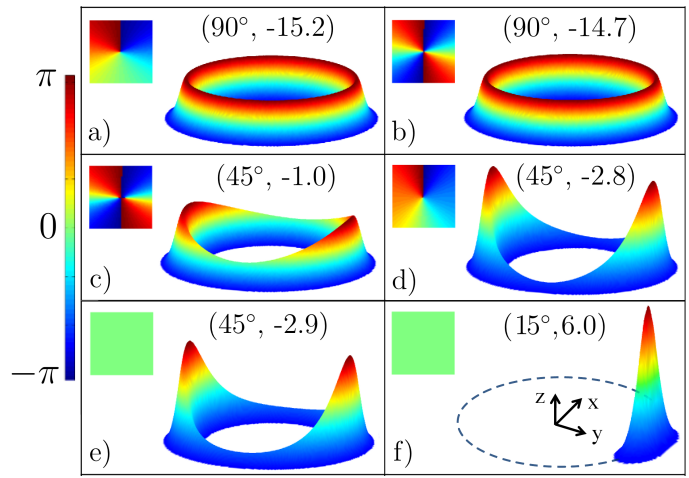}}
\caption{(Color online)
Two-dimensional density distribution of the atoms, 
corresponding to different points $(\Theta,g/g_0)$ 
of the phase diagram of Fig.\,2. The phase of the order 
parameter for each configuration is also shown in the 
insets, with the associated colorbar represented on the 
left of the figure. Panels (a) and (d) correspond to 
metastable states with one unit of circulation, (b) and 
(c) to metastable states with two units of circulation, 
and (e), (f) correspond to current-free states. The dashed 
line in (f) is drawn to guide the eye along the minimum of 
the annular potential.}
\label{fig3}
\end{figure}

In Fig.\,3 we show the two-dimensional density distribution
and the phase of the order parameter corresponding to 
different values of $\Theta$ and $g/g_0$ within the blue 
and the green regions of the phase diagram of Fig.\,2. 
When the dipole moment is oriented perpendicularly to 
the plane of motion of the atoms, the dipolar interaction 
is isotropic and purely repulsive, resembling the contact 
potential, and as a result the density is axially symmetric. 
Panels (a) and (b) show the density for $\Theta = 90^\circ$ 
and $g/g_0 = -15.2$ and $-14.7$, corresponding to metastable 
states with one and two units of circulation, respectively. 
Since the density is axially symmetric, the expectation value 
of the angular momentum per particle for these current-carrying 
states is $L/N = 1$ and 2, respectively.

Panels (c), (d) and (e) show the density for $\Theta=45^\circ$ 
and $g/g_0 = -1.0, -2.8$ and $-2.9$ respectively, corresponding 
to states with $m = 2, 1$, and 0. In this case, the dipolar 
interaction is anisotropic and as a result the cloud concentrates 
around two opposite ends of the annulus, as discussed earlier 
\cite{Aba10}. Notice that the states shown in (c) and (d)
support persistent currents, even though the density is not 
axially symmetric. As a result, $L/N$ does not have an integer 
value and it is equal to $L/N=1.87$, 0.56 in panels (c) and (d), 
respectively.

Finally, in panel (f) we show the density for $\Theta=15^\circ$ 
and $g/g_0=6.0$. In this case there is a spontaneous 
symmetry-breaking due to the dipolar interaction and the density 
localizes in one of the two degenerate minima of the dipolar 
potential, in close analogy to the case of toroidal trapping 
\cite{Aba10}. The system then rotates as a solid body and 
persistent currents are not stable. Indeed, our results indicate 
that the presence of only one density maximum excludes the 
possibility of metastability. 

\section{Summary and overview}

In this study we have considered a dipolar Bose-Einstein
condensate that is confined in an annular potential. The 
atoms are assumed to interact via the dipolar interaction, 
plus an extra short-ranged potential. The dipoles are 
assumed to be aligned along a direction that forms an 
angle $\Theta$ with their plane of motion. Our two basic 
results are the stability of the ground state against 
collapse and the stability of persistent currents as a 
function of the angle $\Theta$ and of the strength of 
the hard-core potential $g$.  

The hard-core and the dipolar interactions combine, 
giving rise to an effective interaction that is spatially 
dependent. Furthermore, this effective interaction depends 
on the angle of orientation of the dipoles, which can be
tuned externally by simply tilting the applied polarizing
field. It is well known that the strength of the contact 
potential may be tuned via the so-called Feshbach resonances 
\cite{Wer05}. In our suggested physical system the tunability 
becomes possible via the change of the orientation of the 
external polarizing field. In addition, in the case of 
Feshbach resonances one may only change the value of the 
scattering length, while here the effective coupling 
is spatially dependent. 

In the phase diagram $g$ versus $\Theta$ we determined two 
boundaries, one delimitating the region where the system 
is stable against collapse, and another one delimitating 
the region where persistent currents are stable. Both phase 
boundaries have a sinusoidal dependence on the angle $\Theta$, 
in agreement with a simple variational model that we have 
presented. As the strength of the dipolar interaction decreases, 
the variation of $g(\Theta)$ decreases and in the limit that it
vanishes, the two phase boundaries which we have found become 
horizontal straight lines, corresponding to the limit of purely 
contact interactions. 

We have seen that metastability of the flow is possible 
even for values of the angle $\Theta$ that give rise to
inhomogeneous density distributions, provided that the 
gas preserves the two-fold symmetry of the dipolar interaction 
and is not concentrated around only one of the minima of the 
dipolar potential. According to our results, provided that
the density is not homogeneous, the presence of two density 
maxima is a necessary, but not a sufficient condition for 
metastability. This situation is somewhat analogous to the 
case of a Bose-Einstein condensate which is trapped in a 
ring potential and interacts via purely contact interactions: 
in this case, if the strength of the contact potential is 
sufficiently large in magnitude and negative, there is a 
spontaneous symmetry breaking of the axial symmetry of the 
Hamiltonian. The ground-state density localizes and it is 
energetically favorable for the system to carry angular 
momentum via solid-body rotation, in which case the system 
does not support persistent currents. 

An appealing property of this system is that one may sweep 
through all the three phases by simply changing the orientation 
angle of the dipoles, even when all the other parameters are kept 
fixed. An interesting potential application of this fact 
might be a ``mesoscopic superfluid ring", i.e., a device that 
is either ``normal" or ``superfluid", depending on the orientation 
of the external polarizing field. Therefore, this problem is 
interesting not only for purely theoretical reasons, but 
also in connection with applications on e.g., precision 
measurements and nanotechnology.

\section{Acknowledgements}

We thank G. Bruun, J. Cremon and J. Smyrnakis for useful 
discussions. This work was financed by the Swedish Research 
Council. This project originated from a collaboration within 
the POLATOM Research Networking Programme of the European 
Science Foundation (ESF).

\appendix*
\section{Evaluation of the dipolar energy}

The interaction between two dipoles with dipole moment 
$\bf d$ located at the positions ${\bf r}$ and ${\bf r}'$
is given by
\begin{eqnarray}
V_{\rm dip, 3D}({\bf r},{\bf r}') =
D^2 \frac {1-3\cos^{2} \theta_{rd}} {|{\bf r} - {\bf r}'|^{3}},
\end{eqnarray}
where $\theta_{rd}$ is the angle between the relative position 
vector ${\bf r} - {\bf r}'$ and ${\bf d}$. The relationship
between $D$ and $d$ is given in Sec.\,II.

First we integrate over the vertical and transverse directions, 
assuming that the order parameter has a product form $\Psi({\bf r}) 
= Z(z) \Phi_{\perp}(s_{\perp}) \Phi_l(s_l)$, with
\begin{eqnarray}
Z(z) = \frac 1 {(\pi a_z^{2})^{1/4}} \, e^{-z^2/(2 a_z^2)},
\end{eqnarray}
\begin{eqnarray}
\Phi_{\perp}(s_{\perp}) = \frac 1 {(\pi d_{\perp}^{2})^{1/4}}
\, e^{-s_{\perp}^2/(2 d_{\perp}^2)}.
\end{eqnarray}
The effective one-dimensional potential between two atoms which 
are located at the points $s_l$ and $s_l'$ along the annulus is
\begin{eqnarray}
V_{\rm dip, 1D}(s_l,s_l') =
\int V_{\rm dip, 3D}({\bf r},{\bf r}')
\, |Z(z)|^2 |\Phi_{\perp}(s_{\perp})|^2
\nonumber \\
\times |Z(z')|^2 |\Phi_{\perp}(s_{\perp}')|^2 \,dz\, ds_{\perp}
\, dz' \, ds_{\perp}'.
\end{eqnarray}
Since the confinement along the $z$ direction is very tight, 
i.e., $a_z \ll d_{\perp}$, we consider for convenience the limit 
$a_z \to 0$. In this case,
\begin{eqnarray}
  \frac 1 {\sqrt{2 \pi} \, a_z} \, e^{-z_r^2/(2 a_z^2)} 
  \to \delta(z_r),
\end{eqnarray}
i.e., we get a delta function in the integration over the variable 
$z_r = z-z'$. In this case $V_{\rm dip, 1D}$ is given by the 
approximate expression
%
\begin{eqnarray}
V_{\rm dip, 1D}(s_l,s_l') = \frac {\sqrt{2 \pi}} {4 \pi}
\frac {D^2} {d_\perp^3} \left[
\frac {\sqrt{\pi}} {4 {\tilde s}^2} U(3/2,0, 2 {\tilde s}^2) +
\right.
\nonumber \\ + \left. \frac 2 3 (1 - 3 \sin^2 a)
e^{{\tilde s}^2} \left[ 2 {\tilde s}^2 K_0({\tilde s}^2) -
(2 {\tilde s}^2-1) K_1({\tilde s}^2) \right] \right],
\nonumber \\
\label{app}
\end{eqnarray}
where $U$ is the confluent hypergeometric function, ${\tilde s} 
\equiv s_{r}/(2 d_{\perp})$, with $s_{r}=s_l - s_l'$ being the 
relative coordinate, and $\sin^2 a \equiv \sin^2 (S/R) \cos^2 
\Theta$. 

Both for small and for large values of ${\tilde s}$ the term 
in the last line of Eq.\,(\ref{app}) is the dominant one. For 
$|s_l - s_l'| \gg d_{\perp}$ Eq.\,(\ref{app}) reproduces the 
correct asymptotic behavior of the dipolar potential,
\begin{eqnarray}
V_{\rm dip, 1D}(s_l,s_l') \approx \frac {D^2} {s_r^3} 
(1 - 3 \sin^2 a).
\label{appp}
\end{eqnarray}
Furthermore, for small values of $\tilde s$, Eq.\,(8) follows.

Assuming that the longitudinal wavefunction is also Gaussian,
\begin{eqnarray}
\Phi_l(s_l) = \frac 1 {(\pi d_l^{2})^{1/4}}
\, e^{-s_l^2/(2 d_l^2)},
\end{eqnarray}
the dipolar energy $E_d$ of the system is given by
\begin{eqnarray}
E_d = \int V_{\rm dip, 1D}(s_l,s_l')
\, |\Phi_l(s_l)|^2 |\Phi_l(s_l')|^2 \, ds_l \, ds_l'.
\end{eqnarray}
Since the dominant contribution to this integral comes from 
small values of $\tilde s$, we use the expansion of
$V_{\rm dip, 1D}(s_l,s_l')$, given by Eq.\,(\ref{vdip1d})
to find that
\begin{eqnarray}
E_d &\approx& \frac {\sqrt{8 \pi}} {3 \pi}
\frac {D^2} {d_{\perp}}
(1 - 3 \sin^2 a) \int \frac 1 {s_r^2}
\, |\Phi_l(s_l)|^2 |\Phi_l(s_l')|^2 \, ds_l \, ds_l' 
\nonumber \\
&=& \frac {4} {3 \pi}
\frac {N D^2} {d_l d_{\perp} a_z}
\left(1 - 3 \sin^2 a \right).
\label{ttrial}
\end{eqnarray} 
In the integration over the relative coordinate $s_r$ 
we have introduced a cut-off, which we have set equal 
to $a_z$. This cut-off comes from the fact that in the 
derivation of the potential $V_{\rm dip, 1D}(s_l,s_l')$ 
we have taken the limit of $a_z \to 0$, which gives a 
delta function in the integration over $z-z'$. As a 
result, in the initial integration, where we have the 
term $(x-x')^2+(y-y')^2+(z-z')^2$ in the denominator, 
we have set $z=z'$. However, in the actual calculation 
$a_z$ is small but finite. To take this into account, 
we have replaced the lower limit of $s_{r}$ in the 
integration over this variable by the ``width" of the 
delta function, or in other words with the average 
value of $\sqrt{\langle (z-z')^2 \rangle}$, which is 
$\sim a_z$.


\begin{thebibliography}{99}

\bibitem{Grie} A. Griesmaier, J. Werner, S. Hensler, J. Stuhler, 
and T. Pfau, Phys. Rev. Lett. {\bf 94}, 160401 (2005).

\bibitem{Fatt} M. Fattori, G. Roati, B. Deissler, C. D'Errico, 
M. Zaccanti, M. Jona-Lasinio, L. Santos, M. Inguscio, and G. Modugno, 
Phys. Rev. Lett. {\bf 101}, 190405 (2008); S. E. Pollack, D. Dries,
M. Junker, Y. P. Chen, T. A. Corcovilos, and R. G. Hulet, ibid. 
{\bf 102}, 090402 (2009).

\bibitem{Veng} M. Vengalattore, S. R. Leslie, J. Guzman, and 
D. M. Stamper-Kurn, Phys. Rev. Lett. {\bf 100}, 170403 (2008).

\bibitem{Lu} M. Lu, S. H. Youn, and B. L. Lev, Phys. Rev. Lett. 
{\bf 104}, 063001 (2010).

\bibitem{Cosp} C. Ospelkaus, S. Ospelkaus, L. Humbert, P. Ernst, 
K. Sengstock, and K. Bongs, Phys. Rev. Lett. {\bf 97}, 120402 (2006).

\bibitem{Sosp} S. Ospelkaus, A. Pe'er, K.-K. Ni, J. J. Zirbel, 
B. Neyenhuis, S. Kotochigova, P. S. Julienne, J. Ye, and D. S. Kin, 
Nat. Phys. {\bf 4}, 622 (2008).

\bibitem{Bar} M. A. Baranov, Phys. Rep. {\bf 464}, 71 (2008).

\bibitem{Lah} T. Lahaye, C. Menotti, L. Santos, M. Lewenstein, 
and T. Pfau, Rep. Prog. Phys. {\bf 72}, 126401 (2009).

\bibitem{Gup} S. Gupta, K. W. Murch, K. L. Moore, T. P. Purdy, 
and D. M. Stamper-Kurn, Phys. Rev. Lett. {\bf 95}, 143201 (2005).

\bibitem{Ols} S. E. Olson, M. L. Terraciano, M. Bashkansky, and
F. K. Fatemi, Phys. Rev. A {\bf 76}, 061404(R) (2007).

\bibitem{Ryu} C. Ryu, M. F. Andersen, P. Clad´e, V. Natarajan, 
K. Helmerson, and W. D. Phillips, Phys. Rev. Lett. {\bf 99}, 260401 
(2007).

\bibitem{Hen} K. Henderson, C. Ryu, C. MacCormick, and M. G. Boshier,
New J. Phys. {\bf 11}, 043030 (2009).

\bibitem{Les} I. Lesanovsky and W. von Klitzing, Phys. Rev. Lett. 
{\bf 99}, 083001 (2007).

\bibitem{Aba10} M. Abad, M. Guilleumas, R. Mayol, M. Pi, and D. M. Jezek,
Phys. Rev. A {\bf 81}, 043619 (2010). 

\bibitem{Aba2} M. Abad, M. Guilleumas, R. Mayol, M. Pi, and D. M. Jezek,
Europh. Phys. Lett. {\bf 94}, 10004 (2011). 

\bibitem{Zol11} S. Z\"ollner, G. M. Bruun, C. J. Pethick, and 
S. M. Reimann, Phys. Rev. Lett. {\bf 107}, 035301 (2011). 

\bibitem{Cre10} J. C. Cremon, G. M. Bruun, and S. M. Reimann, 
Phys. Rev. Lett. {\bf 105}, 255301 (2010).

\bibitem{Chi05} S. A. Chin and E. Krotscheck, Phys. Rev. E {\bf 72},
036705 (2005).

\bibitem{Meystre} O. Dutta, M. J\"a\"askel\"ainen, and P. Meystre,
Phys. Rev. A {\bf 73}, 043610 (2006).

\bibitem{Santos} S. Sinha and L. Santos, Phys. Rev. Lett. {\bf 99},
140406 (2007).

\bibitem{Frank} F. Deuretzbacher, J. C. Cremon, and S. M. Reimann,
Phys. Rev. A {\bf 81}, 063616 (2010).

\bibitem{Bar10} S. Bargi, F. Malet, G. M. Kavoulakis, and S. M. Reimann,
Phys. Rev. A {\bf 82}, 043631 (2010).

\bibitem{U} R. Kanamoto, H. Saito, and M. Ueda, Phys. Rev. A {\bf 68}, 
043619 (2003).

\bibitem{GMK} G. M. Kavoulakis, Phys. Rev. A {\bf 69}, 023613 (2004).

\bibitem{Wer05} J. Werner, A. Griesmaier, S. Hensler, J. Stuhler, 
T. Pfau, A. Simoni, and E. Tiesinga, 
Phys. Rev. Lett. {\bf 94}, 183201 (2005).

\end{thebibliography}
\end{document}